# Optically Encoded Suspension Microarrays: Materials, Design Strategies, and Future Directions for Multiplexed Bioanalysis

*Mikhail Sokolov*[‡], *Irina S. Kriukova*[‡], *Alyona Sukhanova*[‖], *and Igor Nabiev* [‡,‖,*]

[‡] Nano-Photon Research Center, National Research Nuclear University MEPhI (Moscow Engineering Physics Institute), 115409 Moscow, Russia

[‖] Université de Reims Champagne-Ardenne, BioSpecT UR-7504, 51100 Reims, France

* Corresponding author: I. Nabiev, igor.nabiev@univ-reims.fr

## Abstract

Suspension microarrays based on optically encoded microbeads have become one of the most versatile platforms for multiplexed bioanalysis because they combine solution-phase reaction kinetics, flexible assay design, and high-throughput detection. However, despite more than two decades of intense research, no consensus has emerged regarding the optimal strategies for particle encoding, surface functionalization, and signal decoding. Progress has mainly been driven by incremental improvements in individual materials rather than by systematic comparison of competing technological concepts.

This review critically evaluates the main approaches to the fabrication of optically encoded microbeads, including post-synthetic (swelling and layer-by-layer assembly) and *in situ* encoding strategies, and proposes a mechanistic classification of *in situ* methods of particle formation into polymerization-driven and confinement-controlled ones. Instead of comparing the fabrication methods solely in terms of encoding capacity, we assess their relative merits in terms of structural control, code stability, scalability, compatibility with biofunctionalization, and suitability for clinical implementation. We further examine the strengths and limitations of organic fluorophores, aggregation-induced emission luminogens, semiconductor quantum dots, and upconversion nanoparticles and show that no encoding material is universally optimal and that performance is determined by trade-offs between optical properties, manufacturing complexity, and stability in biological media.

We argue that future progress will depend not as much on increasing the theoretical number of optical codes as on improving the code reproducibility, minimizing spectral crosstalk and nonspecific interactions, and employing microfluidic fabrication, antifouling surface chemistry, automated spectral decoding, and artificial intelligence–assisted data analysis. These developments are expected to transform suspension microarrays from multiplexed analytical tools into standardized lab-on-a-microbead platforms suitable for next-generation clinical diagnostics.

## Introduction

Advances in clinical diagnostics critically depend on the development of technologies for highly sensitive, multiplexed analysis. Because most pathological processes are heterogeneous, simultaneous detection of a wide range of biomarkers is required, which makes traditional platforms insufficiently effective, given their high labor intensity and high sample consumption [1, 2]. Therefore, multiplexed technologies have been intensely developed in past decades. Currently, planar (solid-phase) arrays and suspension microarrays are the most popular approaches. Their key operational and analytical characteristics are compared in Table 1.

**Table 1.** Comparison of planar and suspension microarray platforms for multiplexed analysis

| Characteristic | Solid-phase microarrays | Suspension microarrays |
|---|---|---|
| **Probes** | Oligonucleotides, peptides, antibodies | A wide range of biomolecules (antigens, antibodies, other proteins, glycopolymers, oligonucleotides) |
| **Interaction kinetics** | Heterogeneous phase; the reaction kinetics is limited by diffusion to the flat surface | Quasi-homogeneous phase; high mass transfer rate in the solution |
| **Panel tunability** | Low (the arrays have a fixed architecture) | High (the types of microbeads are combined as required) |
| **Multiplexing capacity** | Extremely high for nucleic acids (>$10^5$ ) but limited for proteins (typically <$10^3$) due to cross-reactivity and partial denaturation of the ligands on the flat substrate | High for protein targets (typically $10^1$–$10^2$, potentially up to $10^3$). An optimal balance between maximizing the number of analytes and minimizing the antibody cross-reactivity. |
| **Throughput** | Medium (limited by the time of exposition and scanning) | High (analysis of up to $10^4$ –$10^5$ events per minute) |
| **Reproducibility** | Limited by nonuniform sorption and edge effects | High (due to statistical averaging of data across a large sample of microbeads) |
| **Preservation of** | Proteins may denature during | Protein conformation is preserved in |

| **the native properties of the ligands** | immobilization on a flat surface | a hydrated polymer matrix |
|---|---|---|
| **Signal detection methods** | Laser scanners and high-resolution CCD cameras | Flow cytometry and fluorescence microscopy |

The choice between the planar solid-phase and suspension microarrays is largely determined by the nature of the biospecific ligands. Planar microarrays dominate the field of whole-genome screening: the linear structure and high stability of DNA/RNA molecules allow hundreds of thousands of probes to be placed on a single substrate without loss of functionality. In proteomic analysis, however, rigid immobilization on the surface often leads to protein denaturation and blockage of their active sites.

Suspension microarray platforms avoid this problem due to quasi-homogeneous interaction kinetics and the preservation of native protein conformation in the hydrated polymer matrix of microbeads. Although the multiplexity of suspension systems ($10^2$–$10^3$ analytes) is lower than that of planar arrays, this is sufficient for most clinical diagnostic tasks, where priority is given to the reproducibility and analytical sensitivity of the method rather than the number of ligands.

Suspension microarrays based on optically encoded microbeads have emerged as one of the most technologically mature and scalable platforms for high-throughput multiplexed analysis [3, 4]. Their operating principle consists in the incubation of a biological sample in a medium containing a pool of encoded microbeads whose surface is functionalized with specific ligands to enable selective binding of target analytes (Figure 1).

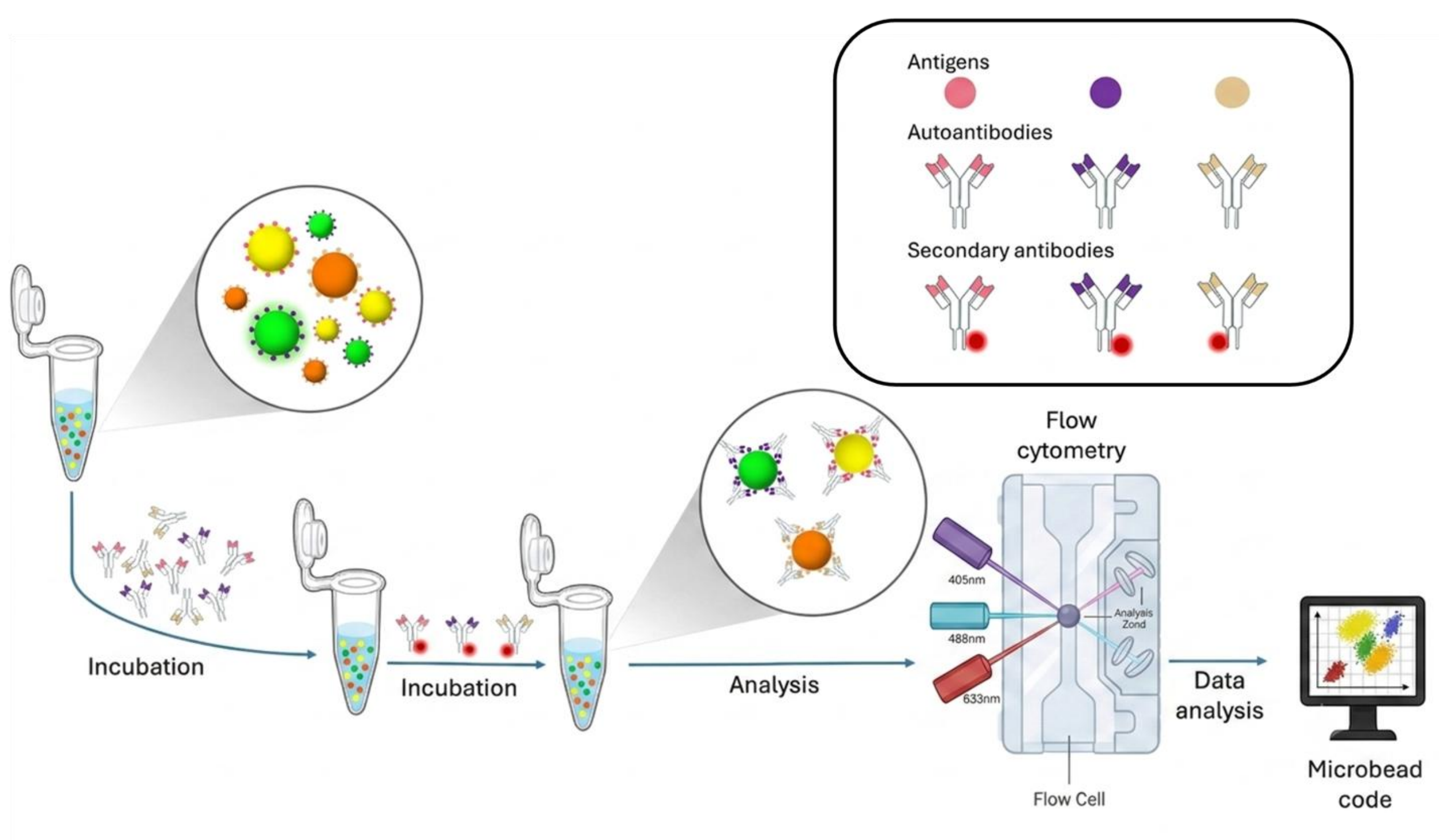

**Figure 1.** Schematics of the operation of suspension microarrays in multiplexed immunoassays.

The analytical signals from microbeads are usually quantitatively detected and decoded using fluorescence microscopy or flow cytometry. Fluorescence microscopy allows for the imaging of individual microbeads with high spatial resolution while simultaneously recording the coding and reporter signals corresponding to the formation of immune complexes on their surface. Flow cytometry provides high-throughput quantitative analysis of microbeads, with tens of thousands of events recorded per minute, and enables simultaneous identification of the particle spectral code and measurement of the intensity of the analytical fluorescence signal. Therefore, it is widely used in clinical diagnosis and multiplexed bioanalyses [5].

Suspension arrays are characterized by a high sensitivity, scalable multiplexing level, and high throughput. Depending on the encoding method, these systems can simultaneously detect from several to hundreds of analytes with a sensitivity comparable to that of classical immunoassays. Note that simultaneous detection reduces the cost of analysis, the cost per marker decreasing with increasing number of targets detected in a single sample.

The advance of nanotechnology has led to the development of new classes of functional luminescent nanoparticles suitable for optical encoding and detection in suspension arrays.

Numerous studies have systematically evaluated their optical characteristics, stability, biocompatibility, and limitations, as well as compared the strategies for encoding microbeads [6–10]. One of the first commercially successful platforms for multiplexed analysis using suspension microarrays was the xMAP system (Luminex), where organic fluorescent dyes were used as encoding labels [11].

The limitations of classical organic dyes, such as spectral overlap, low photostability, and the need for complicated adjustment of the equipment, have spurred the use of alternative materials for the encoding of microbeads. In this context, quantum dots (QDs) have proved promising. They are characterized by a high photostability, narrow emission spectrum, and broad absorption spectrum, which allows particles of different colors to be excited with a single radiation source [12]. However, the excitation of QDs in the short-wavelength spectral region may be accompanied by an increased background signal due to the autofluorescence of biological components. Furthermore, the spectral characteristics of QDs strongly depend on their size, which requires strict control of the synthesis and high monodispersity of the particles.

Upconversion nanoparticles (UCNPs) characterized by anti-Stokes emission are another material used for encoding microbeads. In this case, infrared excitation is used, which significantly reduces the background autofluorescence of biomaterials and minimizes the contribution of Rayleigh scattering, thereby increasing the signal-to-noise ratio [8, 13, 14]. However, UCNPs are characterized by a relatively low quantum yield compared to QDs. Furthermore, the efficiency of upconversion luminescence critically depends on the concentration and uniformity of the nanoparticles. In addition, their integration and stable immobilization in the polymer matrix of microbeads is technologically challenging.

Here we discuss current strategies for increasing the number of available optical codes, improving analytical sensitivity, and refining methods for reading and decoding signals from encoded microbeads in suspension microarrays. In addition, examples of their application to clinical diagnosis are considered.

## Methods for Optical Encoding of Microbeads.

The efficiency of multiplexed suspension platforms is largely determined by the accuracy and reproducibility of the optical encoding, which are crucial factors for unambiguous identification of the microbeads in complex biological environments. The encoding labels form unique codes based on the combination of spectral emission maxima and relative signal intensities that allow differentiation between individual particle populations. The choice of the encoding strategy directly determines the number of distinguishable populations and the compatibility of the platform with the detection system used. The development of versatile and scalable approaches to encoding is a key prerequisite for enhancing the capabilities of multiplexed analysis and efficiency of suspension microarrays in clinical diagnosis.

### *Swelling.*

The method of swelling is a classic strategy for the incorporation of fluorescent labels into polymeric microbeads. It is based on an increase in the free volume of the polymer matrix in an organic solvent. This allows fluorophores or nanoparticles to diffuse into the microbeads, where the labels are fixed upon removal of the solvent. Optical codes are formed by using fluorophores with different spectral characteristics and by varying their concentration, which results in discrete and reproducible populations of microbeads.

Early studies demonstrated the versatility of this approach for various types of fluorophores. Stsiapura et al. [15] showed that the swelling method ensures homogeneous incorporation of CdSe/ZnS QDs into the polymer matrix and the formation of stable fluorescent codes in the microbeads. A similar mechanism underlies the combined swelling–diffusion (CSD) method proposed by Meredith et al. [16], where water-insoluble organic dyes dissolved in tetrahydrofuran (THF) diffuse into swollen polystyrene microbeads and are fixed in them upon removal of the solvent. This approach makes it possible to control the loading of the dye and its distribution within the particles.

Recent modifications of the method aim to increase the incorporation depth and stability of the incorporated nanocomponents. For example, the high-temperature chemical swelling method (HTCSM) enables deeper and more stable incorporation of QDs and magnetic nanoparticles. Upon heating above the glass transition temperature ($T > T$g), the mobility of the polymer chain segments increases, which substantially facilitates the diffusion of nanoparticles into the matrix. Cooling leads to the solidification of the matrix and fixation of the incorporated components, ensuring high signal intensity and stability [17]. The efficiency of incorporation is determined by the ratio of the effective pore size of the polymer network to the hydrodynamic diameter of the incorporated particles. An insufficient free volume limits the diffusion and leads to adsorption of the labels on the surface, whereas excessively large pores increase the risk of partial leaching after solvent removal.

Apart from variation of physical parameters, chemical modification of the polymer matrix allows controlling the thermodynamics of fluorophore distribution and increasing their retention within the particle. For example, the solubility of amphiphilic molecules in the polymer network can be adjusted by varying the polarity of the organic phase using modifiers (such as triethylamine and acetic acid) [18]. As a result, molecules such as fluorescein and tetracycline readily diffuse into the swollen matrix despite their limited hydrophobicity and are retained after the solvent is removed. Copolymerization with functional monomers, such as glycidyl methacrylate (GMA) and tert-butyl methacrylate (tBMA), further enhances the compatibility of the matrix with hydrophilic dyes. The epoxy groups of GMA form polar regions in the polymer and can participate in subsequent covalent conjugation of biomolecules, whereas tBMA provides steric protection and regulates the hydrophobic–hydrophilic balance of the polymer matrix. Combined, these effects extend the range of incorporable fluorophores and increase their retention in the matrix [19]. Optimization of the chemical and physical parameters of the process directly determines the number of optical codes by controlling the absolute and relative concentrations of the incorporated fluorophores.

The swelling method combines operational simplicity with highly homogeneous incorporation of fluorophores throughout the microbead, which helps maintain the monodispersity of the microbeads and increase signal intensity. However, loading of large amounts of the dye may lead to its aggregation, spectral shifts, and reduced reproducibility upon scaling up, because the process is sensitive to the kinetics of swelling and diffusion.

Thus, swelling represents a physically and chemically controlled method of intramatrix optical coding in which the composition of the polymer phase, swelling conditions, and fluorophore properties determine the structure, stability, and reproducibility of the resulting fluorescence codes.

***Layer-by-Layer Adsorption.***

Although the swelling method is effective in forming stable intramatrix codes, alternative approaches, particularly layer-by-layer adsorption (LbL), expand the possibilities for optical encoding of microbeads by providing additional control over the surface structure of the microbeads. The LbL method is based on sequential electrostatic adsorption of oppositely charged polyelectrolytes and nanoparticles to form multilayer shells on the microbead surface [20]. This approach makes it possible to control the layer thickness with a nanometer precision and precisely position the functional components, combining different materials into a single structure. This flexibility makes LbL particularly attractive for obtaining microbeads with desired optical properties.

Early research, such as studies by Rogach et al. [21–23], showed that monochromatic QDs can be durably incorporated into polyelectrolyte layers to form luminescent shells with a predictable nanoparticle distribution, which ensured the reproducibility of their optical properties. This method of incorporation provides a high photostability and precise positioning of QDs, which results in homogeneous, optically active structures suitable for multiplexed analysis.

In later studies [24–26], the LbL method was used to form multilayer structures consisting of alternating layers of poly(allylamine) hydrochloride (PAH) and poly(styrene sulfonate) (PSS) into which CdSe/ZnS QDs were incorporated (Figure 2). This approach enabled the formation of stable populations of optically encoded microbeads, spectrally distinguishable in a flow cytometer upon excitation from a single laser. Practical application of these microbeads for multiplexed detection of cancer biomarkers has been reported [24, 25]. Sokolova et al. [26] obtained six populations of microbeads differing in size and spectral code that were characterized by a high homogeneity, colloidal stability, and differential optical properties. The microbeads were distinctly separated in the standard channels of a flow cytometer. The method ensured reliable fixation of QDs, preservation of reactive groups for subsequent biofunctionalization, and the formation of homogeneous microbead populations with different optical codes, suitable for highly sensitive biosensor systems.

Extending the method to fluorescent magnetic microbeads has made it possible to obtain structures containing QDs and γ-$Fe_2O_3$, thereby combining optical encoding with magnetic functionality. This combination enables highly sensitive multiplexed analysis using a large number of identifiable particle populations [27].

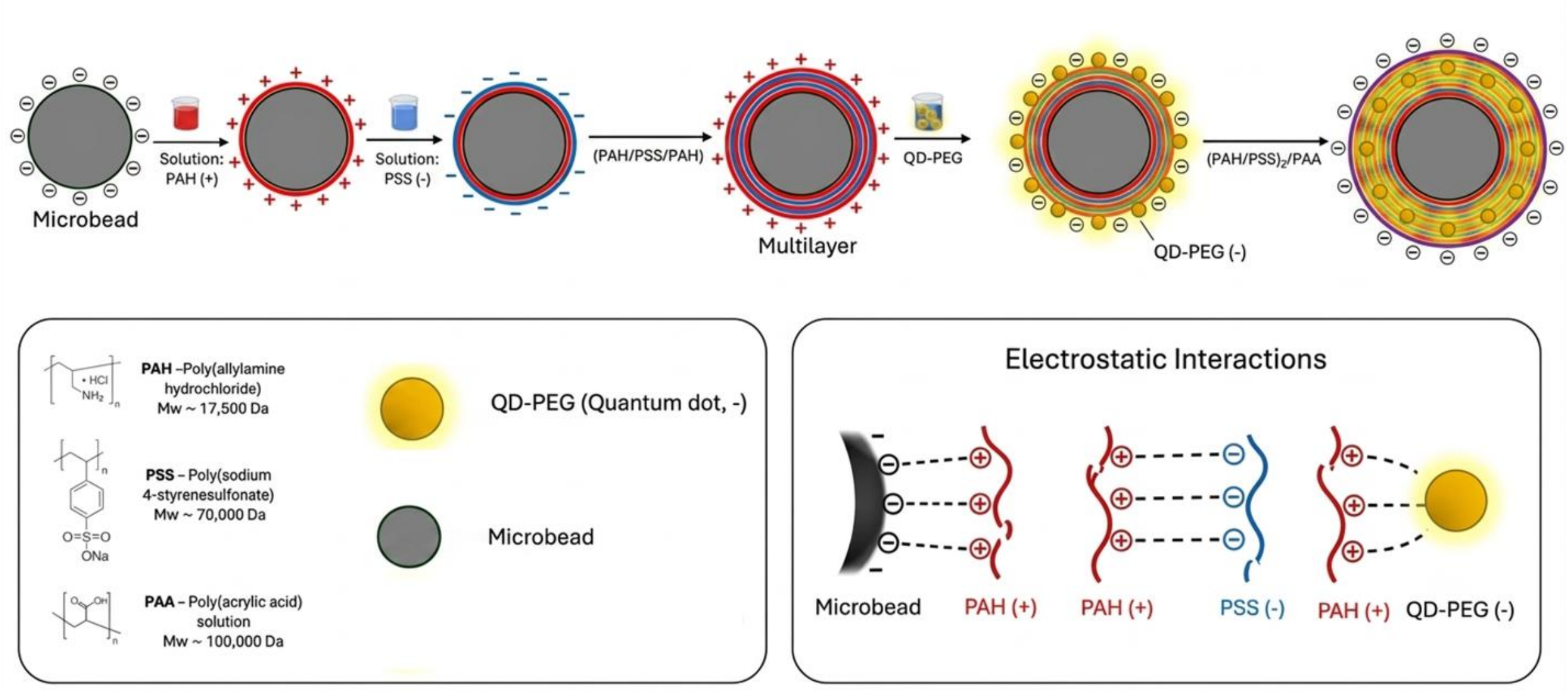

**Figure 2.** Schematics of the layer-by-layer (LbL) assembly on a microbead, with oppositely charged layers deposited sequentially to form a functional shell.

Despite the advantages of the LbL method, the formation of multilayer shells remains labor-intensive. Other drawbacks are that some QDs may be washed off during rinsing and that the reproducibility of the layers directly depends on the stability of the polyelectrolytes and the colloidal stability of the nanoparticles. Nevertheless, the high photostability of the resulting microbeads, the controlled distribution of functional components, and the possibility of their further biofunctionalization make LbL an effective tool for fabricating encoded microbeads and developing next-generation platforms for multiplexed analyses.

## In Situ Encoding as a Promising Strategy for Fabricating Optically Encoded Microbeads

In general, polymer microbeads can be fluorescently encoded using two fundamentally different approaches: post-synthetic (*ex situ*) modification of performed particles and *in situ* incorporation of fluorophores during particle formation. Although both strategies are widely employed in the development of multiplexed biosensing platforms, they differ fundamentally in the physicochemical mechanisms governing the formation of the optical code.

In *ex situ* approaches, fluorescent components are introduced into pre-existing microbeads by means of diffusion-driven loading or surface modification. Examples are the aforementioned swelling method and LbL assembly technique. A major limitation of these strategies is that the spatial distribution of fluorophores is governed by diffusion within a preliminarily formed polymer network. Consequently, precise control over the internal architecture of the particles is difficult, which often results in heterogeneous fluorophore distribution and potential fluorophore leakage in bioanalytical applications.

The current *in situ* encoding strategies, diverse as they may be, can be classified into two main categories according to the parameter that determines the final particle architecture. The first category comprises the methods in which microbeads are formed directly during heterogeneous polymerization. These are the dispersion, emulsion, suspension, miniemulsion, and precipitation polymerization techniques. Although they differ in their experimental implementation, they all share the same fundamental mechanism of particle formation. Initially, the reaction medium consists of a relatively homogeneous mixture of monomers, initiators, stabilizers, and fluorescent components. As polymer chains grow, their solubility decreases, leading to nucleation, phase separation, and subsequent particle growth. Fluorophores are kinetically trapped in the polymer matrix during its formation and become incorporated in the resultant particles. Consequently, the particle size, fluorophore distribution, and final morphology are governed primarily by the nucleation statistics and particle growth kinetics rather than by a predefined geometry.

This mechanism is exemplified by the study by Scholtz et al. [28], who encapsulated CdSe/CdS QDs together with Nile Red into polystyrene microbeads using dispersion polymerization. This study demonstrated that surface modification of QDs with polymer-compatible ligands, combined with the incorporation of divinylbenzene as a crosslinking agent, substantially improved nanoparticle retention in the polymer matrix. However, the radical chemistry underlying heterogeneous polymerization also represents the principal limitation of this strategy. As demonstrated by Sheng et al. [29], free radicals can partly disrupt the native ligand shell of QDs, which activates non-radiative recombination pathways and reduces the photoluminescence quantum yield. Therefore, further development of polymerization-driven approaches largely depends on advances in ligand engineering and selection of the reaction conditions capable of minimizing nanocrystal degradation during synthesis.

Overall, the defining feature of this category is that particle architecture is determined by the kinetics of nucleation and particle growth. Particle formation therefore follows a stochastic process while simultaneously providing excellent scalability and compatibility with large-scale manufacturing.

The second main category comprises the approaches where a specified confined reaction volume is formed. Unlike polymerization-driven synthesis, the particle size and architecture are established before the polymerization or gelation begins. In this case, the governing parameter is the geometry of the isolated microreactor rather than the nucleation kinetics.

These approaches vary in design, but they rely on a common principle: a droplet of a precisely defined volume is formed and subsequently solidified through polymerization, photocuring, or gelation. Each droplet functions as an independent microreactor with a predetermined geometry, while the subsequent curing process merely fixes the predefined structure.

One of the earliest examples of this strategy was reported by Xu and Bakker [30], who developed multicolor QD-encoded polymer microbeads with precisely controlled emitter compositions for optical sensing.

The concept was subsequently extended to a broad range of droplet-templated technologies. Tang et al. [31] employed membrane emulsification to obtain highly monodisperse polylactic acid microbeads suitable for multiplexed immunoassays. Bian et al. [32] fabricated core/shell microbeads with spatially separated encoding and functional domains. Kim J. H. et al. [33] used droplet microfluidics to develop multicompartment hydrogel microbeads in which different compartments contained different optical barcodes and specific capture probes. A related approach was reported by Li et al. [34], who employed microfluidic Pickering emulsions in which nanoparticles stabilized the liquid–liquid interface and simultaneously served as optical encoding elements.

The full potential of confinement-controlled synthesis was demonstrated by Kim B. et al. [35], who developed an on-demand microfluidic platform for producing highly monodisperse QD-encoded microbeads using a T-junction microfluidic chip and *in situ* photopolymerization. Their results showed that predefined droplet formation enabled simultaneous control of the particle size, internal architecture, and optical performance while preserving the luminescence properties of the embedded QDs.

Accordingly, the defining characteristic of this category is that the geometry of the isolated reaction volume serves as the primary governing parameter, whereas polymerization merely solidifies the predetermined particle architecture.

Although both strategies rely on *in situ* fluorophore encapsulation, they are governed by fundamentally different mechanisms of particle formation. In polymerization-driven systems, the particle architecture is determined by nucleation and growth kinetics; in droplet-templated systems, by the geometry of the predefined reaction volume.

These differences determine the practical strengths of each strategy. Polymerization-driven approaches remain the methods of choice for large-scale production of polymer microbeads owing to their simplicity, robustness, and scalability. On the other hand, confinement-controlled platforms offer unprecedented control over particle geometry, fluorophore arrangement, and internal architecture, enabling the fabrication of increasingly sophisticated optically encoded carriers for multiplexed bioanalysis.

These approaches are not competing fabrication routes but instead represent two complementary paradigms of *in situ* particle formation. The mechanistic perspective suggested here provides a unified framework for classifying the existing *in situ* encoding strategies independently of their experimental implementation and establishes a conceptual basis for the rational design of next-generation optically encoded polymer microbeads.

Thus, swelling-based methods, LbL assembly, and *in situ* particle formation strategies constitute a complementary toolkit for optical encoding of microparticles. The swelling approaches are characterized by their operational simplicity and enable efficient loading of fluorophores into preformed polymer matrices. LbL techniques allow precise control over the spatial localization of encoding elements within nanoscale layers and provide versatile routes for surface functionalization. In contrast, *in situ* formation methods—from bulk heterogeneous polymerization to droplet-based microfluidic templating—enable coupling of optical encoding directly to the kinetic trajectory of matrix formation, thereby allowing simultaneous control over particle shape, size, internal architecture, and spatial distribution of functional components. The selection of a specific technological platform is governed by stringent trade-offs between the encoding capacity, fluorescence signal stability, biofunctionalization potential, and manufacturing scalability. There is no universal optimal solution, and the final choice of the encoding strategy should be rationally adapted to the specific analytical requirements of each multiplexed assay.

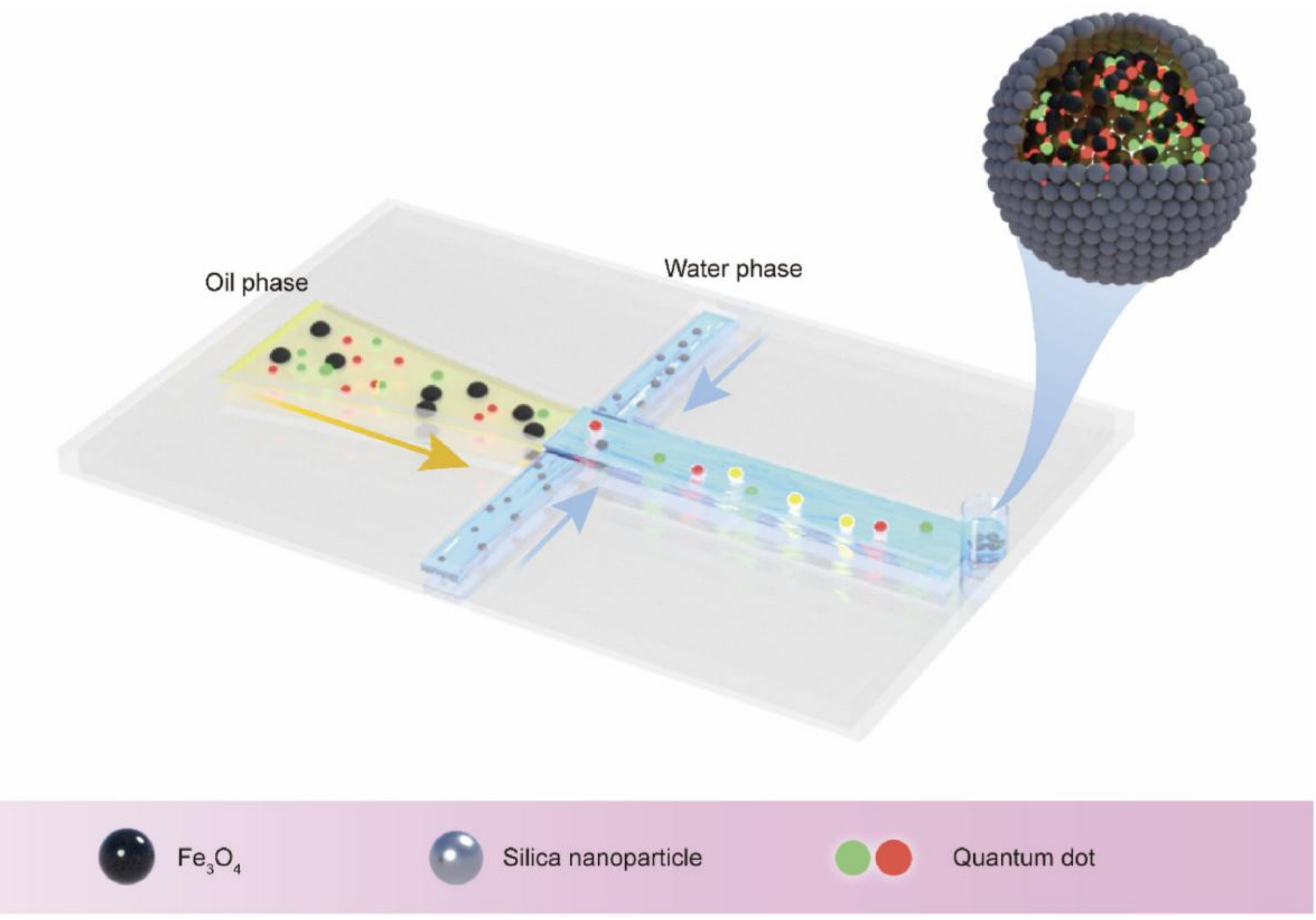

**Figure 3.** Schematics of the synthesis of magnetic fluorescent microbeads using a microfluidic device. Reproduced with permission from Li, Z. et al. Ref. [34] © Elsevier B.V. 2023.

## Materials for the Optical Encoding of Microbeads.

The main classes of materials used for encoding microbeads include organic fluorescent dyes, QDs, and UCNPs, which differ in their spectral properties, photostability, excitation mechanisms, and encoding capacity.

### *Organic Fluorescent Dyes.*

Organic fluorescent dyes have historically been the first and most widely used materials for the optical encoding of microbeads in multiplexed suspension arrays. A classic example is  the xMAP technology (Luminex), which uses polystyrene microbeads about 5.5 μm in diameter containing two spectrally distinct organic dyes. For obtaining unique optical codes, each dye is used at several concentrations, and distinguishable microbead populations are formed by combining different spectral characteristics and luminescence intensities [36, 37]. This allows simultaneous identification of the microbead type and quantity and makes it possible to analyze as many as hundreds of biomarkers within a single measurement [38–41].

In parallel with xMAP, alternative commercial platforms have been developed that offer simpler solutions for multiplexed analysis. For example, the BD Cytometric Bead Array (Becton Dickinson Biosciences) uses ~7.5 μm microbeads encoded with varying concentrations of organic dyes [42, 43]. Unlike xMAP, this system is designed for standard flow cytometers with a 488 nm laser and 576 and 670 nm detection channels, which ensures compatibility with FACScan, FACSCalibur, and similar devices. Although the number of parallel analyses is typically limited (fewer than ten) [44–48], the platform is still an effective tool for immunoassays with simultaneous measurement of several parameters, ensuring convenience and reproducibility of experiments.

Organic fluorophores can be introduced into the polymer matrix during the polymerization stage or incorporated into preformed particles via swelling, which ensures uniform fluorophore distribution and stable optical properties [28, 49]. Sun et al. (2024) proposed a strategy for encoding polystyrene microbeads by combining spectrally distinguishable organic fluorophores and varying their intensity to form unique optical codes suitable for flow cytometry [50]. Another study explored approaches to controlling the ratio between the dye and PEG ligands to form several discrete intensity levels, thereby increasing the number of distinguishable optical codes [51].

However, traditional organic dyes suffer from some serious limitations that complicate the detection and increase the cost of analysis:

- A relatively low photostability.
- Aggregation-caused quenching of photoluminescence.

- Spectral overlap limiting the number of distinguishable codes.
- The need for several excitation sources.

In recent years, aggregation-induced emission luminogens (AIEgens) have emerged as a promising material for encoding microbeads. Unlike classical organic fluorophores, which lose brightness upon aggregation, the luminescence of AIEgens is enhanced upon aggregation due to the restriction of molecular rotational mobility, which reduces energy dissipation and increases the fluorescence quantum yield. AIEgens are also characterized by a large Stokes shift and a high photostability compared to traditional organic dyes [52–54]. Wu et al. [52] demonstrated a suspension array containing four types of AIEgens. They varied the emission wavelength and intensity to generate 30 distinguishable optical codes, which enabled simultaneous detection of five allergens with a high sensitivity (Figure 4).

Despite the advantages of AIEgens, precise control of aggregation is required to achieve their maximum possible fluorescence intensity, because excessively large or unstable aggregates cause light scattering and signal heterogeneity. In addition, large aggregates may precipitate or affect the behavior of microbeads in flow cytometry.

Although AIEgens have better photonic properties compared to classical organic dyes, their use in multiplexed analyses requires careful optimization of the synthesis, aggregation conditions, and compatibility with biological media.

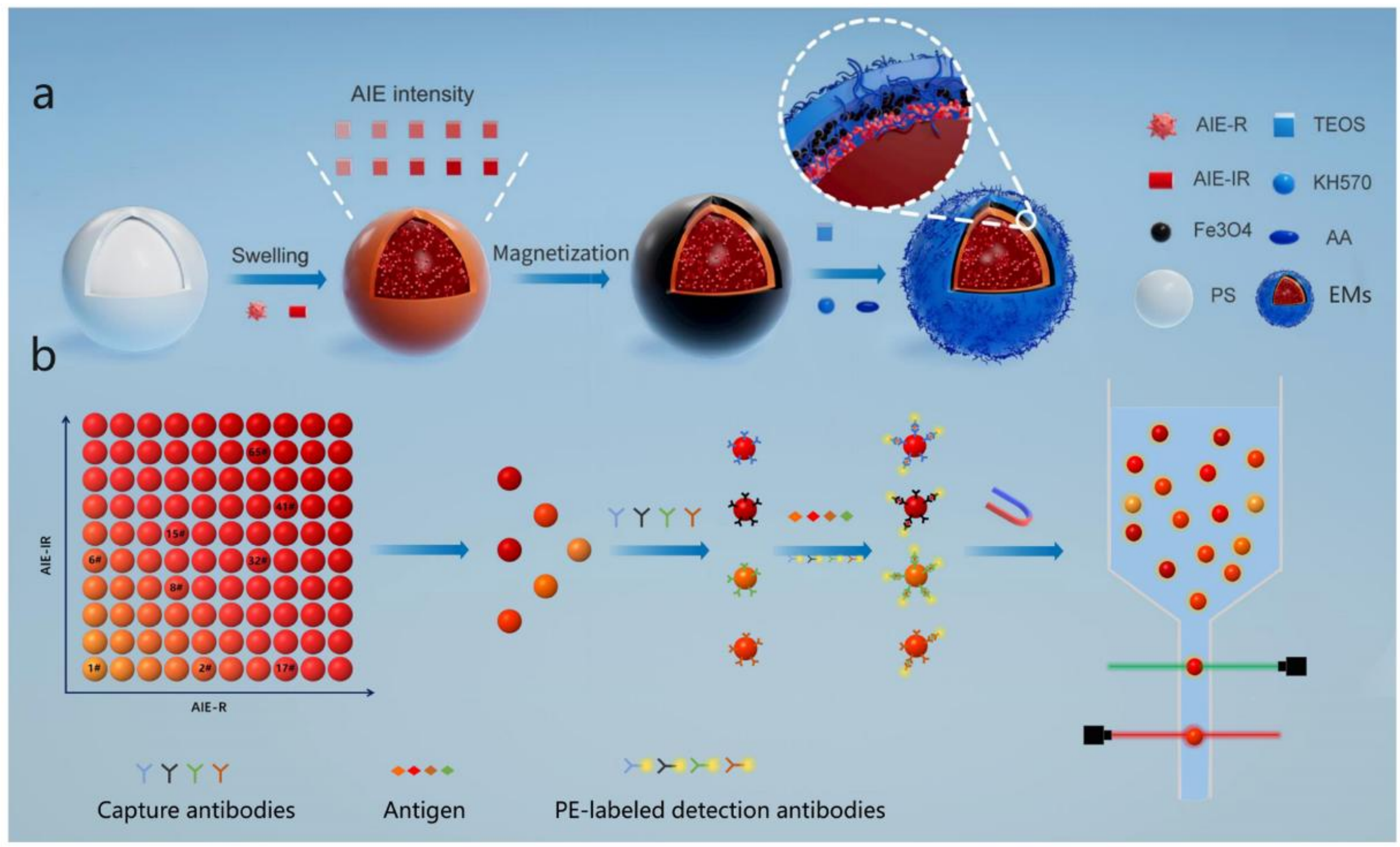

**Figure 4.** Schematics of (a) the formation of encoded microbeads (EMs) and (b) the strategy for combined fluorescence encoding using aggregation-induced emission (AIE) luminogens. Reproduced with permission from Hu, Y. et al. Ref. [52] © Elsevier B.V. 2025.

Although organic fluorescent dyes are widely used in traditional multiplexed systems, their spectral limitations and insufficient photostability have prompted the search for more effective materials. QDs have become the preferred choice for encoding microbeads due to their broad emission spectrum, high photostability, and signal reproducibility, which make it possible to obtain discrete and reliable optical codes.

## *Quantum Dots.*

Although organic fluorescent dyes are widely used in traditional multiplexed systems, their spectral limitations and insufficient photostability have prompted the search for more effective materials. Today, fluorescent nanocrystals or QDs have become the preferred choice for the optical encoding of microbeads. They are characterized by unique properties resulting from quantum size effects [55, 56, 12], including broad absorption and narrow, symmetrical emission spectrum, high photostability, signal reproducibility, and a high quantum yield. These properties enable obtaining discrete and reliable optical codes ensuring unambiguous differentiation of the particles and make QDs a suitable material for encoding microbeads to be used in high-throughput flow cytometry analysis [57, 58, 24].

Quantum dots provide a significantly higher encoding capacity than organic dyes. Theoretically, the maximum possible number of unique codes $C$ is determined as

$$C = N^m - 1,$$

where $N$ is the number of intensity levels and $m$ is the number of colors.

By combining different luminescence wavelengths and intensity levels, it is possible to obtain thousands of unique codes, with all QDs excited from a single light source, which simplifies and accelerates the detection [12].

In 2001, Han et al. [55] incorporated CdSe/ZnS QDs of various sizes into polymer microbeads (1–2 μm) at precisely controlled ratios. The spatial separation of QDs in the microbeads minimized the energy transfer between fluorophores, enhancing the accuracy of identification. This study laid the foundation for the use of QDs as labels for encoding microbeads (Figure 5).

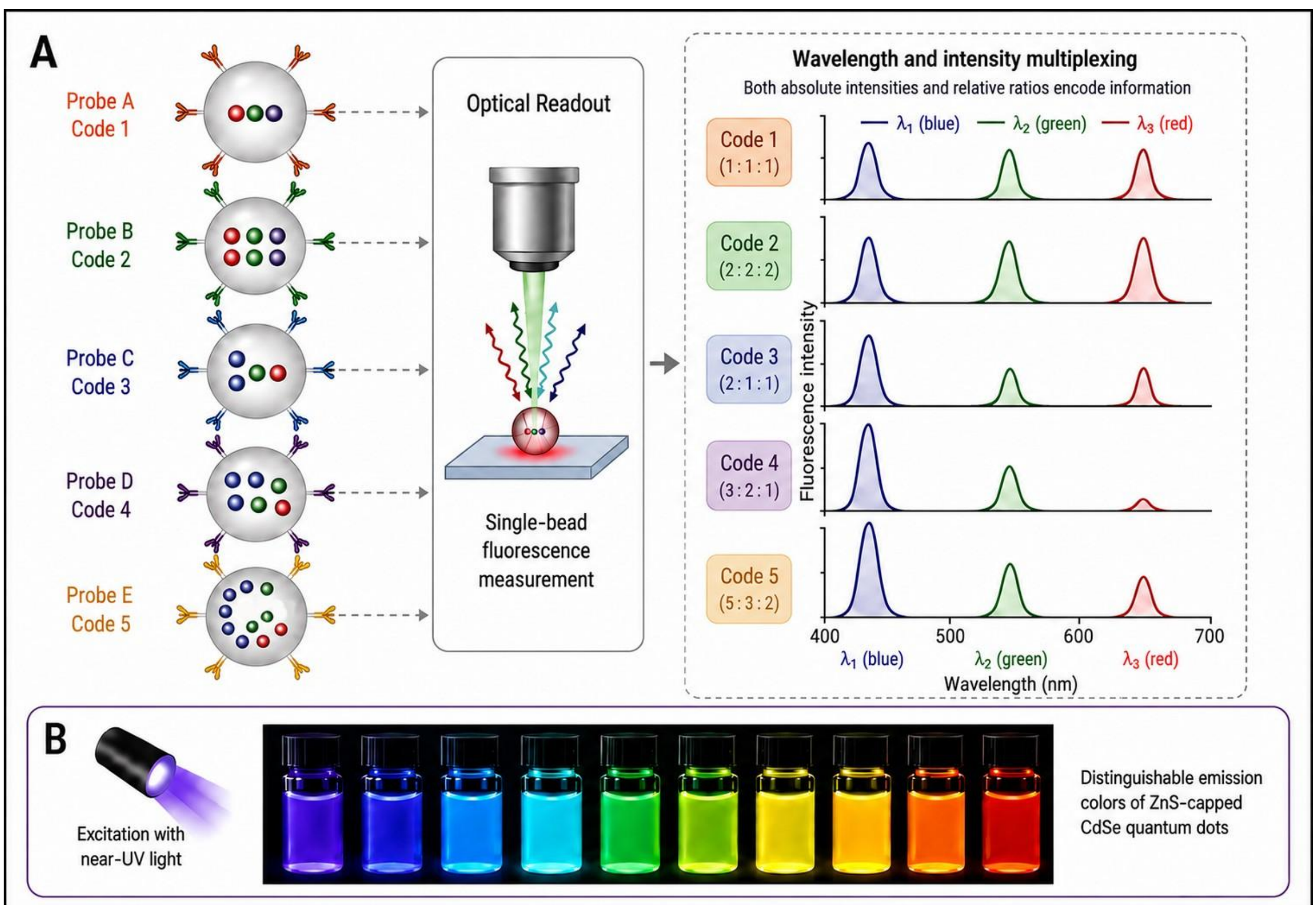

**Figure 5.** (A) Schematic illustration of optical coding based on wavelength and intensity multiplexing. Large spheres represent polymer microbeads, in which small colored spheres (multicolor quantum dots) are embedded according to predetermined intensity ratios. Molecular probes (A–E) are attached to the bead surface for biological binding and recognition, such as DNA–DNA hybridization and antibody–antigen/ligand–receptor interactions. The numbers of colored spheres (red, green, and blue) do not represent individual QDs, but are used to illustrate the fluorescence intensity levels. Optical readout is accomplished by measuring the fluorescence spectra of single beads. Both absolute intensities and relative intensity ratios at different wavelengths are used for coding purposes; for example (1:1:1) (2:2:2), and (2:1:1) are distinguishable codes. (B) Ten distinguishable emission colors of ZnS-capped CdSe QDs excited with a near-UV lamp.

The next step was to enhance the brightness and uniformity of the signals. Gao and Nie [59] demonstrated that mesoporous polystyrene microbeads encoded with QDs exhibited higher intensity and stability compared to non-porous particles. Combining multiple QDs in a single microbead made it possible to identify several fluorescent codes using flow cytometry at a rate as high as 1000 particles per second.

To expand the coding capacity, QDs with a large Stokes shift are used, such as four-armed CdSe/CdS QDs. Wu et al. [60] compiled a three-dimensional library of microbeads with 144 distinguishable codes by combining four-armed CdSe/CdS and CdSe/ZnS QDs and controlling the size of the microbeads. This approach minimized the reabsorption and Förster resonance energy transfer (FRET) between fluorophores, ensuring precise signal separation. The four-armed CdSe/CdS QDs are optimal for reducing FRET and improving the accuracy of multiplexed analyses.

Magnetic nanoparticles are used to expand the functionality and selectively isolate microbeads from the suspension. Early studies showed that direct contact between QDs and the magnetic core partly reduced fluorescence intensity, decreasing the encoding accuracy [61]. The latest solutions minimize this effect by spatially separating QDs and magnetic nanoparticles.

Sha et al. [62] have developed double-encoded microbeads in which QDs are encapsulated in the shell, and a magnetic core is located at the center of the particle. The microfluidic technology ensures precise control of the core size, and the spatial separation of the encapsulated QDs minimizes fluorescence self-quenching effects, ensuring the stability of the optical signal. The spatial separation of the QDs and the magnetic core is a key factor in maintaining bright microbead luminescence.

*Limitations and ways to overcome them*

Apart from their numerous advantages, QDs have also a number of limitations:

- Light reabsorption and FRET between the fluorophores.
- Fluorescence quenching upon contact with magnetic particles.
- The influence of biological media, which may reduce signal intensity and cause false-positive results.

There are advanced approaches that can help minimize these drawbacks:

- The use of four-armed CdSe/CdS QDs to reduce FRET and reabsorption.

• Control of the spatial arrangement of QDs and magnetic nanoparticles in the microbeads.

• The use of anti-fouling coatings and stable composite microbeads to enhance stability in biological media.

Thus, QDs combine a high photostability, narrow emission spectra, and the possibility of excitation from a single light source, which makes them particularly valuable for high-throughput multiplexed platforms. At the same time, in dealing with complex biological systems, it is necessary to take into account signal crosstalk, minimize the fluorescence background, and optimize readout conditions for ensuring accurate and reproducible quantification of each analyte.

### *Upconversion Nanoparticles.*

Upconversion nanoparticles are lanthanide-doped nanocrystals capable of absorbing low-energy infrared radiation and emitting higher-energy light in the visible spectrum (anti-Stokes emission) [63, 64]. The absorption of infrared light by biological media is minimal, which reduces background fluorescence and enhances the detection sensitivity and accuracy. UCNPs are characterized by a high photostability and narrow emission spectra with a minimal spectral overlap, which makes them particularly suitable for multiplexed analysis [9].

The emergence of UCNPs has opened up new possibilities for the development of multifunctional microbeads. Zhang et al. [65] developed UCNP-conjugated magnetic microbeads (UCNMMs) for multiplexed immunoassays. Simultaneous encapsulation of UCNPs and magnetic $Fe_3O_4$ nanoparticles ensured their uniform distribution within the microbeads. Excitation with near-infrared light (980 nm) resulted in visible light emission without overlap of reporter signals, and the magnetic particles enabled rapid recovery of the microbeads from the suspension. The system exhibited high sensitivity and signal stability during prolonged irradiation, which confirmed the potential of UCNPs for microbead encoding [65] (Figure 6).

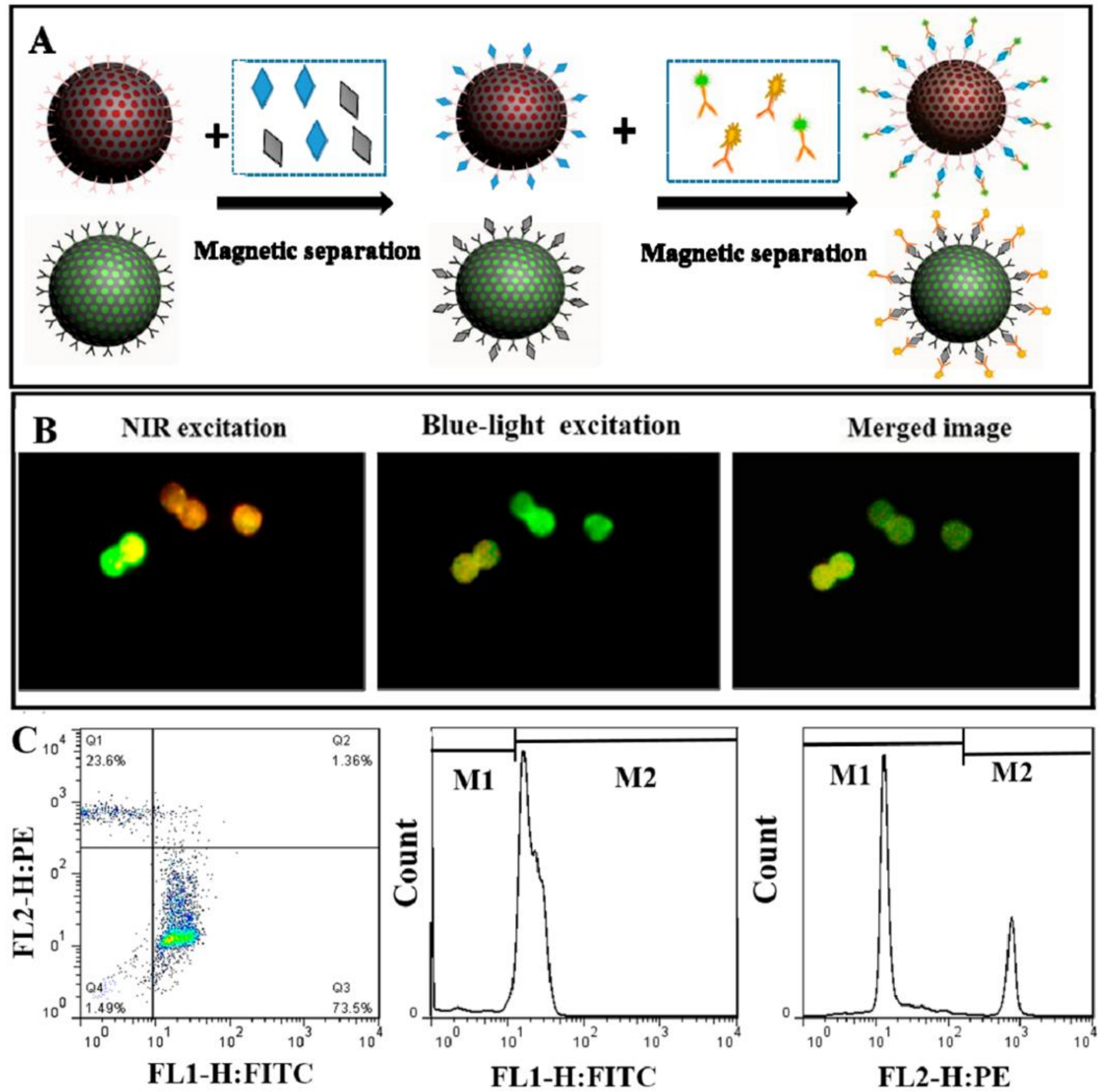

**Figure 6.** (A) Schematics of a double sandwich immunoassay using UCNMM, displaying orange and green fluorescence. (B) Fluorescent images of the two-antibody sandwich immunoassay. (C) Results of the two-antibody sandwich immunoassay analyzed by flow cytometry. Reproduced with permission from Zhang, Y. et al. Ref. [65] © American Chemical Society 2015.

In subsequent studies [66], UCNP populations with different color ratios were encapsulated in polyethylene glycol diacrylate hydrogel microbeads to form multidimensional spectral codes. Microfluidic methods ensured particle homogeneity and uniform nanoparticle distribution. Determination of the ratio between the intensities of different channels allowed simultaneous qualitative and quantitative detection of codes in a single sample, which demonstrated the potential of UCNPs for high-precision multiplexed analysis [66].

A recent study has shown that monochromatic UCNP-encoded microbeads can be used for high-precision spectral identification in multiplexed analysis using flow cytometry. The integration of several types of UCNPs and optimization of the synthesis conditions minimized the signal overlap and increased the measurement reproducibility. The results of these studies highlight the potential of UCNPs for microbead encoding, where accurate spectral identification is combined with high photostability and biocompatibility. This makes it possible to analyze complex biological systems and develop multiplexed diagnostic platforms [67].

Despite their numerous advantages, UCNPs have limitations related to their optical properties. The emission peaks depend on the composition of the lanthanide dopants, which limits spectral tuning. The quantum yield of typical UCNPs rarely exceeds 1% and is the lower the smaller is the particle, which limits the brightness of the encoded microbeads [68].

The fluorescence properties of the particles also depend on the dopant concentration: at low concentrations, upconversion is insufficient, whereas at concentrations exceeding the optimal level, concentration quenching occurs. This is particularly critical for small nanoparticles, where most of the dopants are located near the surface, and energy is lost without the generation of photons [69]. To minimize these effects, core/shell structures are used and centers with a long lifetime are introduced, which increases the photostability and enhances the signal intensity.

UCNP-encoded microbeads have unique optical properties that minimize crosstalk and ensure a high reproducibility of measurements, unattainable with organic dyes or traditional QDs. To employ the full potential of these systems, it is necessary to address the challenges of minimizing concentration quenching and energy losses at the particle surface. Overcoming these limitations will pave the way for the next generation of highly sensitive and scalable multiplexed platforms suitable for fundamental research and clinical diagnosis.

## Methods of Microbead Surface Functionalization.

Efficient surface functionalization of microbeads is a critical step in the development of multiplexed biosensors. The particle surface should allow for the immobilization of capture biomolecules, such as antibodies, antigens, or oligonucleotides, while preserving their biological activity. The sensitivity, specificity, and reproducibility of the assay largely depend on the chemical nature of the surface, the density of functional groups, and the immobilization conditions.

***Methods of Covalent Functionalization of Microbeads Using 1-Ethyl-3-(3-Dimethylaminopropyl) Carbodiimide and N-Hydroxysuccinimide***

Early approaches to the conjugation of biomolecules with microbeads employed electrostatic binding. This strategy ensures rapid immobilization of proteins on the particle surface; however, the bonds are insufficiently stable, and the method is poorly scalable. This limits its applications in modern multiplexed systems [70].

Today, carbodiimide chemistry using 1-ethyl-3-(3-dimethylaminopropyl) carbodiimide (EDC) and N-hydroxysuccinimide (NHS) is the main approach for covalent functionalization of microbeads. The principle of the method is based on the activation of carboxyl groups on the microbead surface using EDC. This results in an unstable intermediate complex, which is then stabilized with NHS, a reactive NHS ester being formed. This ester readily reacts with the amino groups of proteins or peptides, forming strong amide bonds and ensuring reliable conjugation of the components without additional chemical linkers [71].

Several studies have demonstrated that the covalent functionalization of microbeads using EDC/NHS is a versatile approach. In systems with optically encoded microbeads, antibodies are firmly immobilized on the particle surface, ensuring the stability of the complexes [24, 70, 72, 73]. As a result, the conjugation efficiency remains constant across repeated experiments, which is crucial for the reproducibility of the results of multiplexed analysis. A study on covalent binding of antigens using EDC/NHS has demonstrated that the approach is applicable to various biomolecules. However, the efficiency of the EDC/NHS approach strongly depends on the reaction conditions. The activation of carboxyl groups is optimal in a slightly acidic medium (pH ~4.5–6.0); at neutral or alkaline pH, the intermediate complex rapidly hydrolyzes, reducing the reaction yield and the density of biomolecule immobilization. In addition, nonspecific binding of amino groups at protein active sites may somewhat reduce the biological activity of antibodies or peptides. Excessive amounts of reagents sometimes cause microbead aggregation and impair the stability of fluorescent labels. Despite these limitations, carbodiimide chemistry remains one of the most reliable and adaptable methods of covalent bioconjugation, particularly in multiplexed applications, where a high density of functional group and stability of complexes are crucial.

***Strain-Promoted Azide–Alkyne Cycloaddition***

Strain-promoted azide–alkyne cycloaddition (SPAAC) is one of the most efficient methods for covalent immobilization of biomolecules on the microbead surface. Unlike the classical Cu(I)-catalyzed azide–alkyne reaction (CuAAC), SPAAC does not require a metal catalyst, which makes it biocompatible and suitable for proteins, antibodies, and nucleic acids [75, 76]. The reaction is based on the spontaneous cycloaddition of strained cycloalkynes, such as dibenzocyclooctyne (DBCO), to azide groups to form a stable triazole link in an aqueous medium at physiological pH and temperature.

Azidated polymer coatings, such as MCP-6 coating for silicon microbeads or PEG coating for dextran particles, are used for functionalizing optically encoded microbeads [52, 77, 78]. These coatings form a hydrophilic antifouling layer that increases the density and orientation stability of the conjugation of capture molecules and reduce the level of nonspecific binding. DBCO conjugates of oligonucleotides, streptavidin, and antibodies spontaneously react with azide groups on the microbead surface in an aqueous medium under physiological conditions, the functionality and biological activity of the biomolecules remaining intact [51, 78].

SPAAC provides high immobilization density (~5 × $10^6$ molecules per microbead), reproducibility, and binding selectivity, which makes it particularly useful for multiplexed suspension biosensors with a low cross-linking rate and a high sensitivity [75–78].

Briefly, SPAAC has the following advantages:

- No need for metals, which minimizes the toxicity and protein denaturation.
- High reaction rates ensuring rapid binding even at low concentrations of the reagents.
- Compatibility with a wide range of biomolecules, including antibodies, peptides, and oligonucleotides.

However, there are some limitations as well:

- Azide groups should be attached to the microbead surface, which adds one stage to the procedure.
- DBCO conjugates cost more than the classical activators (EDC and NHS).
- The configuration of the coating and size of the microbeads may affect the immobilization density; hence, the conditions of the microbead surface functionalization should be optimized for each particular system.

## Methods of Detection and Analysis

One of the key stages of analysis using optically encoded suspension microarrays is signal decoding, which involves identification of the optical code of each microbead and simultaneous measurement of the analytical signal reflecting the concentration of the corresponding biomarker. For this purpose, flow cytometry and optical imaging methods are widely used which ensure high-throughput quantitative analysis of a large number of particles.

### *Flow Cytometry*

Flow cytometry is a highly sensitive method for analyzing particles in a liquid suspension where they move in a single-file, hydrodynamically focused flow while being irradiated by laser beams. The hydrodynamic focusing is achieved by injecting the sample into a sheath fluid flow, which forms a narrow and stable stream of particles ensuring reproducible detection of signals from individual microbeads with minimal mutual interference.

State-of-the-art flow cytometers are equipped with multiple excitation lasers and a complex optical system based on dichroic mirrors and bandpass filters that enables simultaneous signal detection in 18 fluorescence channels [79–81]. Focusing the laser on a small area around the particle reduces the background signal, and highly sensitive photomultipliers or photodiodes ensure the detection of weak fluorescence responses, which is crucial for detecting small amounts of biomarkers. Parameters of light scattering are used to estimate the particle size and morphology, which also helps in particle identification.

An example of multiplexed analysis is a study [24] that used microbeads encoded with QDs emitting at 515 and 581 nm. Simultaneous detection of signals from the encoded microbeads and the fluorescent reporter (streptavidin–PE-Cy5) bound to target molecules ensured reliable differentiation between particle populations and quantitative assessment of the target binding efficiency. The narrow emission spectra and high photostability of the QDs ensured reproducible code reading under the conditions of a fast particle flow, which is a key requirement for multiplexed suspension assays.

Despite its obvious advantages, the method also has limitations. The sophisticated equipment and the need for precise calibration of multiple channels restrict implementation of the method to well-equipped laboratories. Furthermore, complex multiplexed panels with a large number of encoding and reporter signals require accurate spectral compensation.

Nevertheless, the combination of a high throughput and the possibility of simultaneous reading of the encoding and reporter signals make flow cytometry a major technological platform for detecting optically encoded microbeads in advanced biochemical and diagnostic systems [24, 25, 49, 82–85].

***Alternative Imaging Systems.***

Methods for imaging optically encoded microbeads take an intermediate position between standard plate-based immunoassays and high-throughput flow cytometry. Typically, these approaches use optical or fluorescence microscopy with multichannel signal detection, the particles being identified on the basis of the spectral and intensity characteristics of individual microbeads. Imaging systems can be relatively simple, e.g., optical microscopes with UV or laser excitation and CCD detectors. The resulting color images of the microbeads are decoded using specialized image analysis software [12].

The main difference of the microscopy method from flow cytometry is the shift in focus from maximizing the throughput to obtaining the maximum amount of information on each individual particle. Microscopy enables simultaneous recording of the encoding and reporter signals, as well as the spatial distribution of the immune complex components, providing visual confirmation of specific binding. This makes these methods promising for multiplexed bioanalyses, where control of cross-interactions and accurate interpretation of signals at the level of individual microbeads are crucial. For example, it has been shown [12] that the use of submicron microbeads encoded with QDs makes it possible to detect individual particles and estimate the correlation of fluorescence intensity with the expression rate of membrane proteins, including P-glycoprotein, in multidrug-resistant MCF7 adenocarcinoma cells.

Advanced platforms employ multidimensional decoding that takes into account the spectral characteristics, size, concentration, and spatial distribution of the microbeads. The implementation of machine learning algorithms makes it possible to scale up visual decoding from tens to millions of particles while retaining high accuracy and reproducibility. An example is a digital immunological platform developed by Zhao et al. [86], where microbead parameters are automatically extracted from microscopic images, ensuring highly sensitive, multiplexed biomarker detection across a wide dynamic range of concentrations.

Microscopic digital immunoassays, in which microbeads serve both as places where the immune complexes are formed and as signal reporters, constitute a separate class. In the fluorescent microsphere-based digital immunoassay, after the formation of biotinylated immune complexes, the microbeads are selectively captured by streptavidin-coated magnetic particles and detected by means of wide-field fluorescence microscopy. This approach provides ultra-low detection thresholds (tens of picograms per milliliter or lower) [87]. Similar concepts are used in platforms with core/shell/shell magnetic microbeads for multiplexed analysis of microRNAs [88], as well as in multi-LOCI systems for parallel detection of cytokines [89].

Despite their advantages, microscopy decoding methods have limitations: a relatively low throughput compared to flow cytometry systems, dependence of sensitivity on the optical configuration, and high demands for image analysis algorithms. These limitations are driving the development of hybrid solutions that combine microscopy and automated decoding.

In the future, these platforms may evolve from research tools into versatile multiplexed diagnostic systems, particularly in the cases where visual verification of binding, ultra-sensitive digital readout, and adaptability to new biomarkers are required.

## Representative Clinical and Bioanalytical Applications.

Suspension microarrays based on optically encoded microbeads constitute an efficient platform for simultaneous detection of multiple biomarkers in small biological samples. Owing to their capacity for distinguishing unique optical codes of particles and recording the associated analytical signals, these systems enable simultaneous quantitative detection of tumor markers, cytokines, autoantibodies, and other biomolecules. This ensures integrated assessment of the body condition, accelerates early diagnosis of diseases, and reduces the amount of biomaterial required for analysis. The multiplexing capabilities of suspension microarrays increase the accuracy and reproducibility of the results, optimize the consumption of reagents and laboratory resources, and offer opportunities for monitoring the pathological processes. These platforms are particularly valuable in studying complex diseases with multiple biomarkers of oncological, infectious, and autoimmune pathologies [8, 12].

Representative clinical applications are summarized in Table 2 and the analytical performance of representative optically encoded suspension microarrays is presented in Table 3. These studies collectively demonstrate that optically encoded suspension microarrays provide analytical sensitivities comparable to established immunoassays while enabling simultaneous determination of multiple biomarkers from limited sample volumes.

**Table 2. Representative Clinical Applications of Optically Encoded Suspension Microarrays**

| Application | Target analytes | Encoding material | Detection platform | Multiplexing (used / barcode capacity) | Ref. |
|---|---|---|---|---|---|
| **Prostate cancer** | | | | | |
| Prostate cancer screening | fPSA, tPSA | CdSe/ZnS quantum dots | Multiplex flow cytometry–based sandwich immunoassay | 2 / NR | Brazhnik et al., 2015 [25] |
| Prostate cancer biomarker discrimination | f-PSA, c-PSA | Dual-color CdSe/ZnS quantum dots | Magnetic bead–based multiplex flow cytometry immunoassay | 2 / NR | Min et al., 2022 [83] |
| **Other cancers** | | | | | |
| Lung cancer biomarker profiling | AMBP, PRDX2, PARK7 | CdSe/ZnS quantum dots | Multiplex flow cytometry–based sandwich immunoassay | 3 / NR | Bilan et al., 2017 [24] |
| Tumor marker detection | CA125, CA199, CA724, CEA | CdSe/ZnS quantum dots | Suspension array flow cytometry immunoassay | 4 / 9 | Tang et al., 2022 [31] |
| Tumor marker panel | CA125, CA199, AFP, NSE | CdSe/ZnS quantum dots and $Fe_3O_4$ magnetic nanoparticles | Magnetic fluorescence suspension array | 4 / 30 | Zhang et al., 2023 [89] |
| **Infectious diseases** | | | | | |
| COVID-19 immune profiling | Anti-S1 IgG, anti-N IgG, anti-RBD IgG, IL-6, IL-1β, PCT, CRP | Organic dyes | Microfluidic fluorescence immunoassay with RCA amplification | 7 / 44 | Lin et al., 2022 [90] |
| **Autoimmune diseases** | | | | | |
| Systemic autoimmune disease profiling | Anti-SSA/Ro60, anti-CENP B, anti-RNP70, anti-Scl-70, anti-histones | Organic dyes | Multiplex flow cytometry immunoassay with MFI readout | 5 / NR | Chauhan et al., 2022 [91] |
| Primary biliary cholangitis | AMA-M2, anti-gp210, anti-sp100 | Organic dyes | Multiplex bead–based flow fluorescent immunoassay | 3 / NR | Wang et al., 2024 [92] |
| **Allergy** | | | | | |
| Multiplex allergen-specific IgE | Pollen, milk, peanut, egg white, house dust mite sIgE | AIE luminogen (AIEgen) | Multiplex flow cytometry fluorescence immunoassay | 5 / 30 | Wu et al., 2019 [54] |

| profiling | | | | | |
|---|---|---|---|---|---|
| Multiplex allergen-specific IgE profiling | Artemisia, cat hair, dog hair, *D. pteronyssinus*, *D. farinae*, *Humulus* sIgE | Organic dyes | Multiplex fluorescence immunoassay | 6 / NR | Sun et al., 2024 [50] |

**Table 3. Analytical Performance of Representative Optically Encoded Suspension Assays**

| Ref. | Biomarker panel | Sample | Assay time | Analytical range | LOD/LOQ | Clinical performance | Advantages | Limitations |
|---|---|---|---|---|---|---|---|---|
| Brazhnik et al., 2015 [25] | fPSA, tPSA | Serum (50 µL; *n* = 68) | ~110 min | fPSA: 0.07–10 ng/mL; tPSA: 0.5–60 ng/mL | fPSA: 0.067 ng/mL; tPSA: 0.12 ng/mL | High correlation with ELISA; reproducible quantification (CV = 2.3–2.7%) | High sensitivity; low sample consumption; simultaneous detection of PSA isoformrs | Need for a flow cytometer; limited multiplexing (2 analytes); inter-assay variability up to 30% |
| Min et al., 2022 [83] | f-PSA, c-PSA | Serum (tPSA 4–10 ng/mL) | ~3 h | c-PSA: 0.2–50 ng/mL; f-PSA: 0.64–80 ng/mL | f-PSA: 22 pg/mL; c-PSA: 45 pg/mL | Strong quantitative agreement ($R^2$=0.997); enables PSA isoform discrimination | Single-laser excitation; stable reporters (>6 months); magnetic separation reduces bead loss | Longer assay time; need for a flow cytometer |
| Bilan et al., 2017 [24] | AMBP, PRDX2, PARK7 | BALF (42 patients, 10 controls) | NR | NR | LoQ: 7–10 ng/mL | Combined biomarker panel exhibits a high diagnostic performance (AUC = 0.992) | Multiplex biomarker detection in a single sample; comparable with Luminex xMAP® | Small cohort; only preclinical validation; high QD cost |
| Tang et al., 2022 [31] | CA125, CA199, CA724, CEA | Serum-like standards | NR | Wide linear range (NR) | NR | Feasibility for multiplex tumor marker detection | High-throughput detection; low CV (<10%); low background fluorescence | Limited barcode capacity (9 codes); Cd toxicity concerns; lack |

| | | | | | | | | of clinical validation |
|---|---|---|---|---|---|---|---|---|
| Zhang et al., 2023 [89] | CA125, CA199, AFP, NSE | NR | Magnetic separation ~3 min; total assay time NR | NR | CA125: 0.89 KU/L; CA199: 0.72 KU/L; AFP: 0.05 ng/mL; NSE: 0.15 ng/mL | Multiplex tumor marker detection with a high encoding capacity | High barcode capacity (30 codes); magnetic enrichment; low nonspecific binding | Complex nanoparticle synthesis; Cd toxicity concerns; limited clinical validation |
| Guo et al., 2020 [88] | IFN-γ, IL-4, IL-10, IL-17A | PBMC supernatants (15 donors) | No-wash mix-and-measure (time NR) | ~5 orders of magnitude (IFN-γ) | NR | High-precision cytokine quantification (CV 5.0–10.3%) | No-wash workflow; no hook effect; multiplex cytokine profiling | Need for a specialized imaging system; complex barcode fabrication; limited biological validation |
| Lin et al., 2022 [90] | Anti-N IgG, anti-S1 IgG, anti-RBD IgG, IL-6, IL-1β, PCT, CRP | Serum/plasma; whole blood compatible | NR | ~10 orders of magnitude | IL-6: 0.54 pg/mL; IL-1β: 6.5 pg/mL; PCT: 1.15 ng/mL; CRP: 0.35 μg/mL | Integrated immune and inflammatory profiling for COVID-19 assessment | High multiplexing capacity; enhanced sensitivity (30–50×) ensured by RCA; microfluidic integration | Assay time not reported; limited precision data |
| Chauhan et al., 2022 [91] | Anti-SSA/Ro60, anti-CENP B, anti-RNP70, anti-Scl-70, anti-histones | Serum (*n* = 150) | ~90 min | Linear dilution range up to 1:1600 | NR | Sensitivity: 89.5–95.5%; specificity: 88.9–100% vs IIFT | Objective quantitative MFI readout; faster than IIFT; good reproducibility (CV = 7.6–10.0%) | Limited antigen panel; need for antigen storage at −80°C; need for further validation |
| Wu et | Allergen- | Serum (*n* = 95) | ~2 h | 0.001–100 | 0.004–0.008 | Strong agreement with | High signal | Lower coding |

| al., 2019 [54] | specific IgE (5 allergens) | | | IU/mL | IU/mL | ImmunoCAP ($R^2$≈0.99); sensitivity 82–90%; specificity 92–98% | amplification; single-laser detection; quantitative allergen profiling | capacity compared with QD systems; small cohort |
|---|---|---|---|---|---|---|---|---|
| Sun et al., 2024 [50] | Allergen-specific IgE (6 allergens) | Serum (*n* = 202) | ~70 min | NR | NR | Strong agreement with ImmunoCAP/ELISA (*r* = 0.85–0.97) | Reduced serum volume; high reproducibility (CV<10%) | LOD not reported; correlation varies among allergens; mixed reference methods |
| Wang et al., 2024 [92] | AMA-M2, anti-gp210, anti-sp100 | Serum (238 PBC, 81 AIH, 62 SLE, 118 controls) | NR | Positivity threshold >20 AU | NR | Sensitivity of combined antibody detection up to 98.3%; agreement with immunoblot of 87.4–95.4% | Quantitative autoimmune serology; improved diagnostic sensitivity using combined markers | Single-center cohort; anti-sp100 alone showed lower performance (AUC 0.68) |

*NR, not reported

Rather than reviewing every reported application individually, we discuss below only representative examples illustrating the principal analytical advantages and remaining limitations of the different optical encoding strategies.

Recent studies show the widespread use of suspension-based platforms based on optically encoded microbeads in cancer diagnosis. For example, suspension systems of microbeads encoded with QDs have been used for simultaneous detection of free and bound forms of prostate-specific antigen (fPSA and tPSA, respectively) [25, 84]. These studies have demonstrated high sensitivity and reproducibility of fPSA and tPSA analyses in a single serum sample (Figure 7).

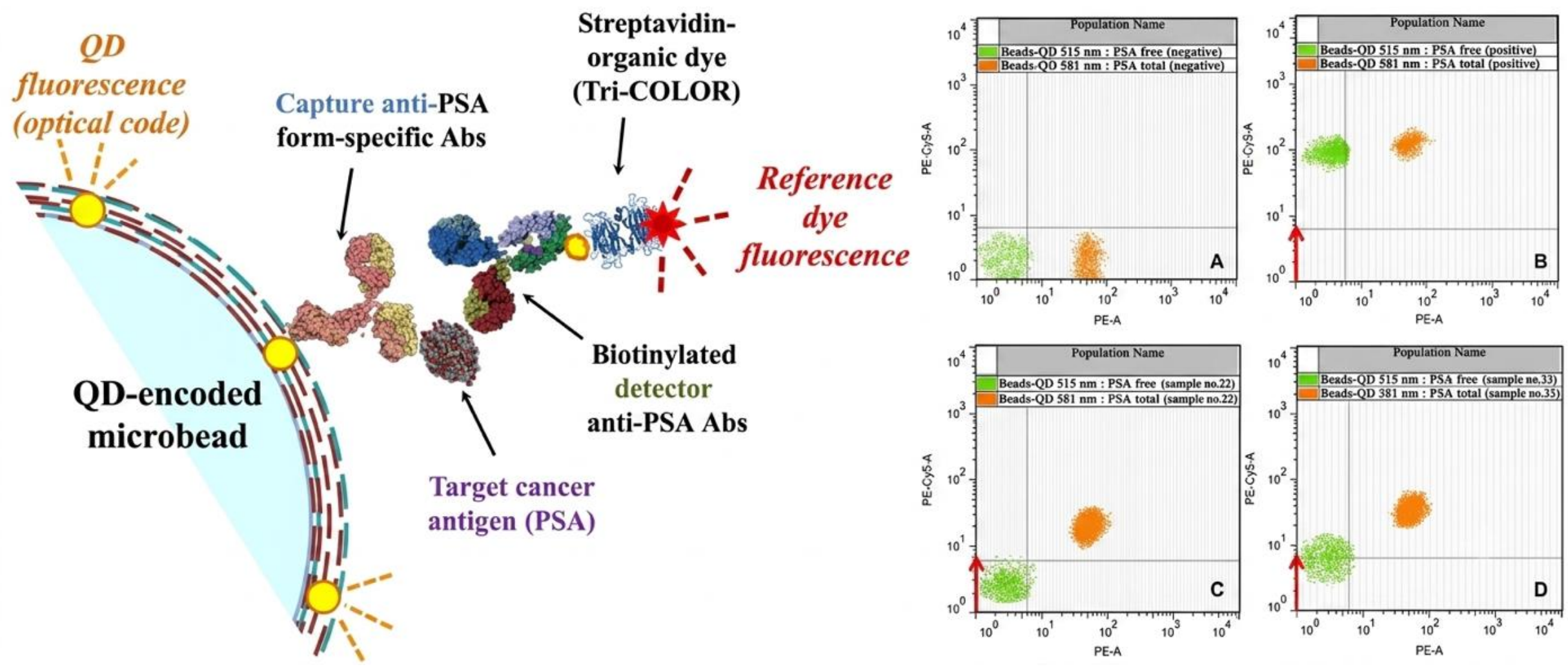

**Figure 7.** Schematics of the lab-on-a-bead system for the detection of prostate-specific antigen (PSA). Reproduced with permission from Brazhnik, K., et al. Ref. [25]. © 2015 Elsevier.

Microbeads encoded with QDs have been used for multiplexed measurement of the levels of cancer biomarkers, including AMBP, PRDX2, and PARK7, the results being comparable to those obtained using the xMAP platform (Luminex) [24]. PLA–MA microbeads encoded with CdSe/ZnS QDs have been used for simultaneous quantitative detection of CA125, CA199, CA724, and CEA [90].

Optically encoded magnetic microbeads for parallel detection of CA125, CA199, AFP, and NSE have been developed [91]. They combine optical identification with magnetic separation to enhance the sensitivity and reduce the background signal.

In the diagnosis of infectious and inflammatory diseases, suspension-based systems ensure simultaneous detection of multiple pathogens and key immune markers, including cytokines, antibodies, and viral antigens, which is critical when symptoms overlap. Platforms based on optically encoded microbeads provide a high sensitivity, a low background noise, and reproducible results [66, 89, 92].

In the field of autoimmune diseases, multiplexed systems enable simultaneous detection of sets of autoantibodies, overcoming the limitations of traditional methods detecting individual antibodies. Chauhan et al. [93] have developed a platform for quantitative detection of five clinically important autoantibodies (SS-A, Ro60, CENP B, RNP70, and Scl70) associated with systemic autoimmune diseases.

Wu et al. [54] have developed a multiplexed system for the diagnosis of allergies based on optically encoded microbeads detecting five allergens (*Artemisia*, milk, peanut, egg protein, and house dust mite). The system is characterized by high stability and reproducibility, a clinical sensitivity of 82–90%, and a specificity of 92–98%, which makes it comparable with ImmunoCAP and confirms the suitability of such platforms for highly sensitive multiplexed assays.

Sun et al. [50] proposed a multiplexed fluorescence immunoassay for the quantitative determination of six allergen-specific IgE (sIgE) antibodies against the most common inhalant allergens: *Artemisia*, cat and dog hair, *Dermatophagoides pteronyssinus*, *D. farinae*, and *Humulus*. The platform performs quantitative analysis within about 70 min, exhibits a high reproducibility (CV < 10%) and a strong correlation with the results obtained using ImmunoCAP and ELISA ($r$ = 0.85–0.97), reduces the required serum volume, and allows for simultaneous testing of multiple allergens in a single sample.

Taken together, available data indicate that suspension-based microarrays using optically encoded microbeads represent a versatile platform for multiplexed diagnostics. They provide high sensitivity and reproducibility, a wide range of measurable biomarker concentrations, and an accuracy comparable to that of traditional methods, while significantly reducing the required sample volume and analysis time. The growing optical code libraries make it possible to simultaneously detect dozens of markers, including tumor, infectious, and autoimmune ones. This offers opportunities for early diagnosis, integrated serological profiling, and therapeutic monitoring of various diseases. The multiplexed format preserves a high level of measurement accuracy and does not compromise analytical performance. The platforms are compatible with the current laboratory equipment, which facilitates their implementation in clinical practice.

## Conclusion and Perspectives

Suspension microarrays have evolved from proof-of-concept multiplexed assays into one of the most mature analytical technologies for simultaneous detection of multiple biomarkers. Nevertheless, analysis of the available literature demonstrates that the field is still dominated by optimization of individual technological components rather than integrated engineering of complete analytical systems. Increasing the number of distinguishable optical codes has frequently been presented as the main technological advance, although the encoding capacity is rarely the limiting factor in practical clinical diagnostics. Instead, assay reproducibility, robustness in complex biological media, stability of surface functionalization, manufacturing scalability, and compatibility with standard analytical instrumentation ultimately determine clinical applicability.

One of the central conclusions from this review is that there is no single optimal encoding strategy. Post-synthetic methods, including swelling and layer-by-layer assembly, remain attractive because of their simplicity, versatility, and compatibility with a wide range of fluorophores. However, diffusion-controlled incorporation inevitably limits precise control of fluorophore distribution and increases the risk of heterogeneity and label leakage. In contrast, *in situ* encoding methods offer substantially greater control of particle architecture but should not be regarded as a single technological class. In our opinion, they can be divided in two groups, each following a fundamentally different paradigm. In polymerization-driven approaches, the particle architecture is determined by the nucleation and growth kinetics, providing excellent scalability but limited structural control. Confinement-controlled approaches, on the other hand, predetermine the particle geometry before solidification, enabling deterministic control over internal organization, spatial separation of the encoding and sensing domains, and increasingly sophisticated multifunctional microbeads. This mechanistic distinction provides a more rational framework for selecting the fabrication strategies than the conventional classification based solely on experimental methodology.

Critical evaluation of encoding materials also demonstrates that their relative advantages strongly depend on the intended application. Organic fluorophores remain attractive because of their low cost and compatibility with commercial instrumentation, but they are inherently constrained by spectral overlap and photobleaching. Aggregation-induced emission luminogens partly overcome these limitations but require precise control of aggregate formation. Semiconductor quantum dots currently offer the best balance between brightness, photostability, spectral tunability, and encoding capacity, making them the most versatile encoding material for multiplexed bioanalysis. Nevertheless, issues related to fluorescence resonance energy transfer, signal reproducibility in biological environments, and regulatory concerns regarding heavy-metal-containing nanocrystals remain incompletely resolved. Upconversion nanoparticles provide unique advantages in the case of near-infrared excitation by minimizing autofluorescence and spectral crosstalk, but their comparatively low quantum efficiency and demanding synthesis currently restrict widespread analytical implementation. Consequently, the optimal encoding material should be selected according to analytical requirements rather than theoretical encoding capacity.

Similarly, advances in surface functionalization have demonstrated that the density of immobilized capture molecules alone is not an adequate predictor of assay performance. Long-term colloidal stability, controlled molecular orientation, suppression of nonspecific adsorption, and preservation of biological activity increasingly determine the analytical sensitivity and reproducibility. Future biofunctionalization strategies should therefore focus on engineering biologically inert interfaces rather than merely maximizing the immobilization efficiency.

From an analytical perspective, future progress will increasingly depend on integrating particle engineering with instrumentation and computational analysis. Automated spectral unmixing, machine-learning-assisted decoding, multidimensional optical barcoding, and real-time quality control will become as important as advances in the development of fluorescent materials themselves. Likewise, microfluidic fabrication offers unique opportunities to standardize particle production, reduce batch-to-batch variability, and precisely control the spatial organization of functional components, thereby improving reproducibility and facilitating regulatory approval.

In our opinion, the next generation of suspension microarrays should no longer be viewed merely as collections of optically encoded beads, but rather as integrated lab-on-a-microbead analytical systems, in which particle architecture, optical encoding, surface chemistry, fluid handling, signal processing, and data analysis are co-designed as a unified platform. Achieving this level of integration will require close collaboration among materials scientists, analytical chemists, microfluidics engineers, computational scientists, and clinical researchers. This convergence is likely to shift the field from incremental improvements in encoding technologies toward robust, standardized, and clinically translatable multiplexed analytical platforms capable of supporting precision medicine.

## Key messages and authors' perspective

- Encoding capacity is no longer the principal bottleneck.
- Structural control is more important than maximal code number.
- *In situ* methods should be classified according to the particle formation mechanism rather than synthesis protocol.
- Quantum dots currently provide the best compromise between brightness and multiplexing capability.
- UCNPs are the most promising platform for low-background assays but require higher quantum efficiency.
- Antifouling surface chemistry is becoming more important than fluorophore selection.
- AI-assisted spectral decoding will become indispensable for highly multiplexed assays.
- The future belongs to standardized lab-on-a-microbead platforms rather than increasingly complex optical barcodes.

**Author Contributions**

Conceptualization, I.N., A.S.; data curation, M.S., I.S.K.; writing—original draft preparation, M.S.; writing—review and editing, A.S., I.N., I.S.K.; supervision and funding acquisition, I.N. All authors have read and agreed to the published version of the manuscript.

**Funding**

This research was funded by the Russian Science Foundation (RSF), grant no. 26-15-00450 “Nano-Photon” in the part related to the optically encoded microbeads design and synthesis, and by the RSF grant no. 22-75-10103-П “CapCan” in the part related to the optically encoded microbeads functionalization.

**Conflicts of Interest**

The authors declare no conflict of interest.

**Acknowledgements**

We thank Vladimir Ushakov for proofreading the manuscript.

**Declaration on the use of Artificial Intelligence**

The authors used OpenAI ChatGPT solely to improve the linguistic quality and readability of the manuscript and to assist in editing. All scientific analysis, interpretation of the literature, critical evaluation, and conclusions were conceived, verified, and approved by the authors, who assume full responsibility for the content of the manuscript.